\documentclass[preprint,floatfix,showpacs,preprintnumbers, 
               amsmath,amssymb]{revtex4}
\usepackage[dvips]{graphics}
\usepackage{psfrag}
\usepackage{graphicx}
\usepackage{epsf}  
\usepackage{ucs}

\def\k{{\bf k}}

\begin{document}
\title{Co-tunneling assisted sequential tunneling in multi-level quantum dots} 

\author {Jasmin Aghassi}
\author {Matthias H. Hettler$^{\,a)}$}
\affiliation{Forschungszentrum Karlsruhe, Institut f\"ur Nanotechnologie,
76021 Karlsruhe, Germany}
\author {Gerd Sch\"on}
\affiliation{Institut f\"ur Theoretische Festk\"orperphysik and
DFG-Center for Functional Nanostructures (CFN),
Universit\"at Karlsruhe, 76128 Karlsruhe, Germany;\\ 
Forschungszentrum Karlsruhe,
Institut f\"ur Nanotechnologie,76021 Karlsruhe, Germany
}


\date{\today}

\begin{abstract}

We investigate the conductance and zero-frequency shot noise of interacting, multi-level quantum dots coupled to leads.
We observe that co-tunneling assisted sequential tunneling (CAST) processes 
play a dominant role in the transition region from Coulomb blockade to
sequential tunneling. 
We analyze for intermediate coupling strength the dependence of the conductance due to CAST processes on temperature, coupling constant, and gate voltage. 
Remarkably, the width of the CAST transport feature scales only with temperature, 
but not with the coupling constant.
While the onset of inelastic co-tunneling is associated with a super-Poissonian noise, the noise is even stronger above the threshold for CAST processes.

\end{abstract}

\maketitle

{\bf Introduction.}
Quantum dots formed in semiconductor heterostructures have been in the focus of interest for transport studies, because they allow considerable variation and control of their properties. The discrete level structure is directly observable, and the coupling to the electrodes can be tuned.
In this letter we study a multi-level quantum dot in a regime of intermediate coupling strength, where besides elastic and inelastic co-tunneling the  
co-tunneling assisted sequential tunneling (CAST) 
processes~\cite{golovach,franceschi} play a crucial role. 
Transport signatures of CAST-processes were observed in the
differential conductance~\cite{schleser}, and their potential limitations for  applications of devices relying on Coulomb blockade have been discussed. In this article we present, starting 
from a perturbative diagrammatic approach, a detailed analysis of the conductance and 
shot noise~\cite{blanter,thielmann,onac,zhang,zarchin} in the experimentally relevant regime. 
We show that the CAST processes lead to characteristic features inside the Coulomb blockade region, particularly dominant in the shot noise. Different from the line width broadening at the threshold to sequential tunneling, the width of the CAST transport feature scales only with temperature, but not with the coupling constant. 

{\bf Model and technique.}
We consider a quantum dot with two spin-degenerate levels with energies $\epsilon_i$ ($i,j=1,2$) and Coulomb interactions $U$ and $U_{nn}$ for intra and inter-level Coulomb repulsion, respectively, 
\begin{equation}
H_{\rm D} = \sum_{i \sigma}\epsilon_{i} n_{i\sigma} 
 + U\sum_i n_{i\uparrow}n_{i\downarrow}  +  U_{nn}\sum_{i \neq j,\sigma,\sigma'} n_{i\sigma'}n_{j\sigma} \, .
\end{equation}
where $n_{i\sigma}= c^{\dagger}_{i\sigma} c_{i\sigma} $ is the number operator of electrons with spin $\sigma $ of the level $i$. 
The dot is coupled to source and drain reservoirs with a total Hamiltonian 
$H=H_{\rm L}+H_{\rm R}+H_{\rm D}+H_{\rm T,R}+H_{\rm T,L}$.
The reservoirs $H_r$ ($r={\rm L,R}$) are modeled by non-interacting electrons
($a_{\k \sigma r}$) with density of states $\rho_r$,
 and the coupling to the reservoirs is described by the tunneling terms
 $H_{{\rm T},r}$ with matrix elements $t_r$
(chosen independent of spin $\sigma$ and equal for both quantum dot levels). 
Together they define the coupling strengths  $\Gamma_r = 2\pi | t_r|^2 \rho_r$ 
and the total linewidth  $\Gamma = \Gamma_{\rm L} + \Gamma_{\rm R}$.
For transport, we apply a source-drain bias $V_{\rm bias}$ 
symmetrically across the quantum dot.
In three-terminal experiments, with an additional gate voltage $ V_{\rm gate}$
applied to the dot, the energies 
of the dot eigenstates are shifted by 
$eV_{\rm gate} {\rm N}$, where ${\rm N} = \sum_{i \sigma} n_{i\sigma}$.

We evaluate the current $I = \langle \hat{I} \rangle$ and 
(zero-frequency) shot noise power spectrum, $ S = \int_{-\infty}^{\infty} dt 
\langle \delta \hat{I}(t) \delta \hat I(0) 
+ \delta \hat{I}(0) \delta \hat I(t) \rangle $,
with $\delta \hat I(t)=\hat I(t)-\langle \hat I \rangle$. The current is the symmetric combination,
$\hat I = (\hat I_{\rm R} - \hat I_{\rm L})/2$
with  current operators defined by
$\hat{I}_r = -i(e/\hbar) \sum_{i \k \sigma} \left(
 t_r a_{\k \sigma r}^{\dag} c_{i\sigma} -{\rm h.c.}\right)$
for electrons tunneling into/out of the lead $r$.
We proceed within the frame of the real-time diagrammatic technique on the 
Keldysh contour~\cite{diagrams}, which was extended in
Ref.~\cite{thielmann,jasmin-prb,jasmin-3qd} to the description of the noise. The resulting
current and shot noise are expressed in terms of 
transition rate matrices ${\bf W}$ (expressed in the basis of eigenstates of $H_{\rm D}$),
for which analytical expressions up to second order 
in $\Gamma_r$, i.e. including co-tunneling processes, were obtained in
Ref.~\cite{jasmin_thesis}. 
For fixed source-drain bias, the stationary probabilities ${\bf p}^{st}$ to occupy
the quantum dot eigenstates can be determined from the ``master'' equation, 
${\bf W}{\bf p}^{st}=0$. Technical details on the expressions for the current and 
shot noise can be found in Ref.~\cite{thielmann}. Rather than plotting the shot noise 
power spectrum $S$ we will present below figures for the
Fano factor, i.e. ratio between the shot noise and current, $F= S/2e I$. 
For uncorrelated transport events (Poissonian statistics) 
the Fano factor is $F=1$. 
Interactions of electrons on the quantum dot may lead to bunching of electrons 
and super-Poissonian Fano factors, $F>1$. 
In contrast, sub-Poissonian Fano factors, $F < 1$, indicate
anti-bunching, which is typical for non-interacting Fermions~\cite{blanter}. 

{\bf Results.}
Transport in quantum dot systems with a large charging energy and weak coupling $\Gamma$ is typically dominated by sequential processes, in which the quantum dot is charged by a single excess electron at a time. The applied source-drain voltage $V_{\rm bias}$ must provide the gate-voltage-dependent energy $\epsilon_{\rm seq} (V_{\rm gate})$ for this single-electron charging. 
This defines the threshold voltage 
$e V_{\rm seq}(V_{\rm gate}) = 2\epsilon_{\rm seq}(V_{\rm gate})$ for symmetrically applied bias above which (at low $T$) sequential tunneling sets in,
while states with higher excess charges are exponentially suppressed. 
This is the case inside the Coulomb blockade region, which in the 
$V_{\rm gate}-V_{\rm bias}$ plane takes a diamond shape. 
In this regime higher order tunneling processes, such as co-tunneling, 
which transport an electron across the dot 
via the {\it virtual} occupation of  dot states, dominate. 
Elastic co-tunneling processes, which leave the dot in its ground state, are the main transport mechanism at very low bias. 
Inelastic co-tunneling processes set in at a bias $eV_{\rm co}= \epsilon_{\rm co}$,
where $\epsilon_{\rm co}$ is the lowest excitation energy {\it within} the ground state charge sector (i.e., for fixed total charge $\rm N$). 
Note that $V_{\rm co}$ does not depend on the gate voltage.

\begin{figure}
\centerline{\includegraphics[width=12cm]{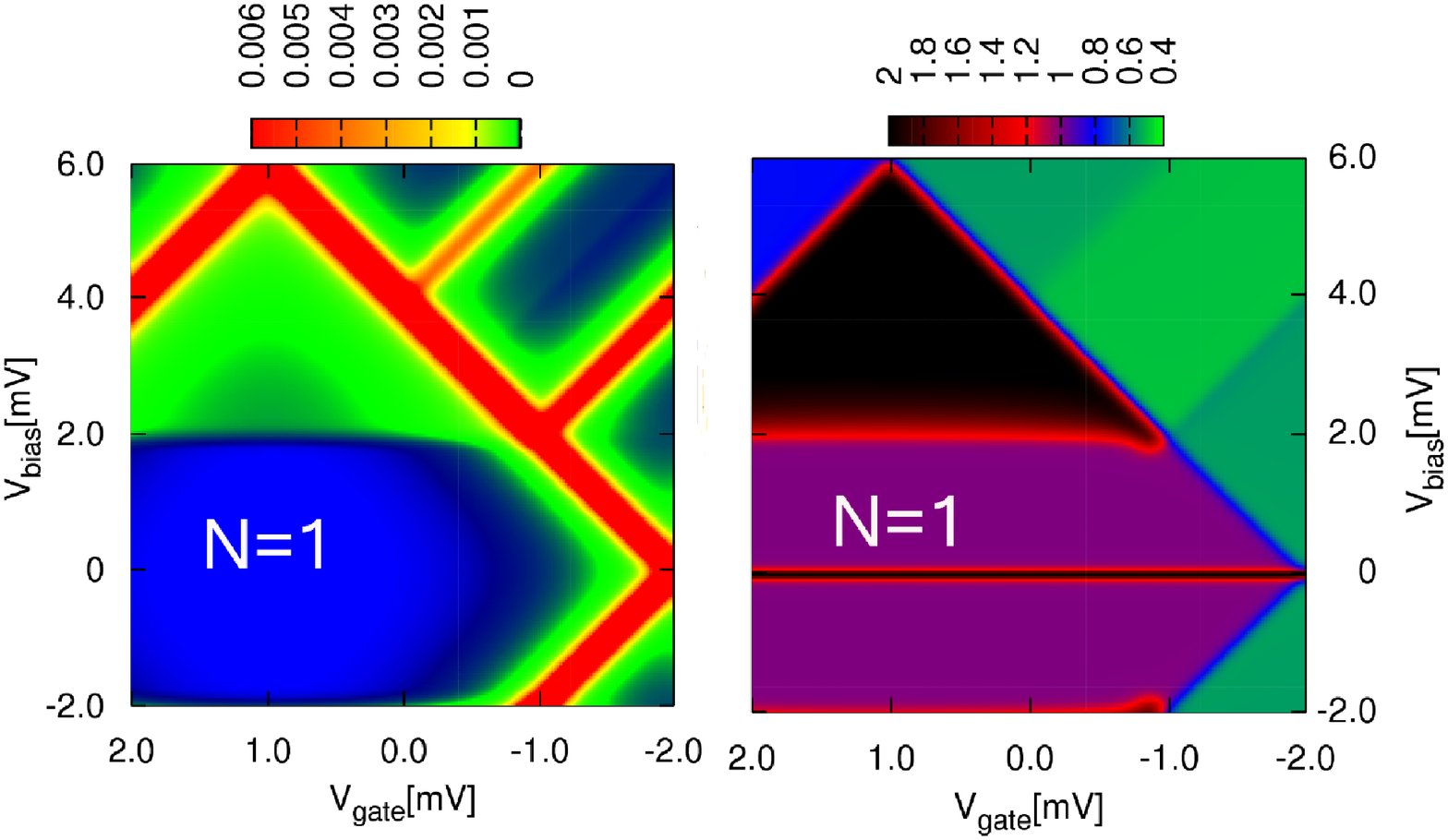}}
 \caption{(Color online) {\it Left panel}: Color-scale plot of the differential
conductance (in units $10\cdot\pi e^2\Gamma/h$) vs. gate voltage ($V_{\rm gate}$) 
and bias voltage ($V_{\rm bias}$). 
The Coulomb diamond edges mark the onset of sequential tunneling. Lines outside the Coulomb diamond correspond to transport via excited states, whereas the horizontal
step inside the diamond is due to the onset of inelastic co-tunneling.
   {\it Right panel}: Color-scale plot of the Fano factor vs. gate and bias voltage. 
Around zero bias voltage the Fano factor diverges due to thermal noise. Poissonian values, $F=1$, indicate that elastic co-tunneling processes dominate at low bias. Super-Poissonian Fano factors occur around the inelastic co-tunneling excitation energy and are reduced to sub-Poissonian values outside the Coulomb diamond.} 
 \label{fig:didv_Fano}
\end{figure}

Throughout this work we choose a temperature of $k_{\rm B}T\le 2\Gamma$ for which a perturbative expansion up to second order 
in the coupling constant $\Gamma$ is appropriate~\cite{thielmann}. 
We begin with parameters $U=20,
U_{nn}=4,\epsilon_1=-4,\epsilon_2=-2,k_{\rm B}T=2\Gamma=0.04$ (all units in meV)
which corresponds to a singly occupied ground state (${\rm N}=1$) at $V_{\rm gate} =0$. 
In the left panel of Fig.~\ref{fig:didv_Fano} we depict the differential conductance $dI/dV$ 
vs.\ the applied bias and gate voltage. The color-scale translates low 
conductance values with dark blue and high conductance values with red. The
Coulomb diamond as well as two lines outside the diamond 
shape which correspond to sequential transport via excited states are
clearly visible. Inside the Coulomb diamond, sequential transport is
exponentially suppressed, while the onset of inelastic co-tunneling 
at $V_{\rm co}$ can be recognized in 
the figure by the horizontal step at
$e V_{\rm co}=2\,{\rm meV}$ (for ${\rm N}=1$ we have $\epsilon_{\rm co}=\epsilon_2-\epsilon_1 =2 {\rm meV}$).
Note, however, that the step height depends on the gate voltage (see discussion below). For a bias below that step only elastic co-tunneling events
contribute to the current, above
both types of co-tunneling, elastic and inelastic, contribute.
Furthermore, since for a ground state charge of ${\rm N}=1$ the low-energy spectrum satisfies the relation $\epsilon_{\rm co} > 2/3 \epsilon_{\rm seq}$ 
for any gate voltage, an inelastic co-tunneling process occupying the upper level can directly be followed by a sequential tunneling process. 
Such transport processes are termed co-tunneling assisted sequential tunneling (CAST) 
processes and will be discussed in more detail below.
Note that second order processes also play a significant role to the conductance at the threshold for sequential tunneling, as they lead to
line width broadening of order $\Gamma$ in addition to the temperature 
broadening already present for pure sequential tunneling processes~\cite{thielmann}.

Next, we discuss the noise properties of the system. 
The right panel of Fig.~\ref{fig:didv_Fano} shows the Fano factor in the same voltage
region as the conductance plot. Outside the Coulomb diamond
the Fano factor drops to sub-Poissonian values, $0.5<F<1$, typical in this region. 
Inside the Coulomb diamond we observe a horizontal, gate-voltage-independent line around zero bias corresponding to the thermal noise in the limit 
$k_{\rm B} T \gg e V_{\rm bias}$. 
For $k_{\rm B}T \ll e V_{\rm bias} < \epsilon_{\rm co}$ the Fano factor becomes 
Poissonian, $F =1$, since only one transport process, namely
elastic co-tunneling, dominates. Increasing the bias further allows for inelastic
co-tunneling beyond $e V_{\rm co}$. 
Now two types of co-tunneling processes with different tunneling rates compete,  
leading to a noisier current with super-Poissonian Fano factors~\cite{thielmann}. 

\begin{figure}
\centerline{\includegraphics[width=12cm]{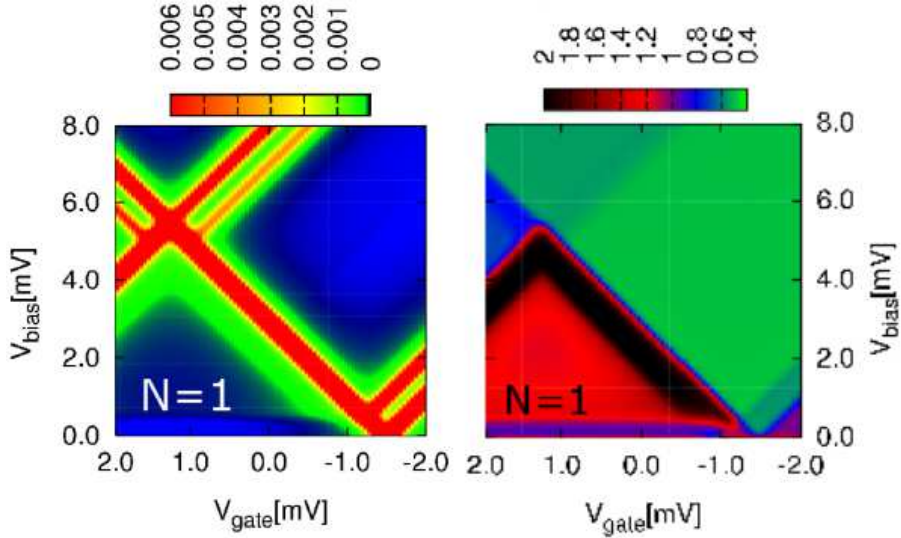}}
\caption{(Color online) {\it Left panel}: Color-scale plot of the differential conductance in the region 
where $ \epsilon_{\rm co}<  \frac{2\epsilon_{\rm seq}}{3}$ (see text). The horizontal step inside the Coulomb diamond
   corresponds to the onset of inelastic co-tunneling whereas in the green, angle shaped region 
additional CAST processes are possible.
{\it Right panel}: Color-scale plot of the Fano factor vs. gate and bias voltage for the
  same parameters as on left panel. The Fano factor becomes super-Poissonian at
  the inelastic co-tunneling excitation energy scale and increases to even higher
  super-Poissonian values when CAST processes set in (edge of the black angle shaped region).
} 
 \label{fig:didv_Fano_triangle}
\end{figure}

From previous theoretical work~\cite{golovach,jasmin_thesis} as well
as experiments~\cite{schleser} it is known that additional structure inside the
Coulomb blockade regime arises if the CAST processes are only possible for a bias 
higher than the inelastic co-tunneling energy. This is the case when the dot spectrum
satisfies the relation $\epsilon_{\rm co}< 2/3 \epsilon_{\rm seq}(V_{\rm gate})$. Using parameters
$U=25,U_{nn}=5,\,\epsilon_1=-4,\,\epsilon_2=-3.5,\,k_{\rm B}T=2\Gamma=0.04$ (all
units in meV), the above relation is satisfied inside the Coulomb diamond
with ground state charge ${\rm N}=1$ for the gate voltages $3.25 > eV_{\rm gate} > -0.75$ 
as depicted in Fig.~\ref{fig:didv_Fano_triangle}.
Now two energy scales are visible. As before, inelastic co-tunneling displays itself in form of a gate-voltage-independent 
conductance step (bright blue to dark blue) at a bias $e V_{\rm co}= \epsilon_{\rm co} = 0.5 {\rm  meV}$. 
The other energy scale corresponds to the threshold for
CAST processes, $e V_{\rm CAST} =2 [\epsilon_{\rm seq}(V_{\rm gate}) - \epsilon_{\rm co}]$, 
visualized by the edge of the bright green ``angular" shape inside the Coulomb diamond. 
These processes are a combination of an inelastic co-tunneling process, 
exciting the quantum dot from the ground state to the first excited state, and a subsequent sequential tunneling processes. The angular edges are parallel to the Coulomb diamond edges since
the gate voltage dependence of the CAST energy scale is the same as for the sequential tunneling threshold 
$e V_{\rm seq}= 2\epsilon_{\rm seq}(V_{\rm gate})$.  For both features, the value of the differential conductance depends on gate voltage.
 
The angular conductance signature of CAST processes has been observed in 
experiments~\cite{schleser}. In that work transport through a
multi-level quantum dot was measured and gate-voltage-dependent lines of the differential conductance inside the Coulomb
diamond were frequently found. It was speculated that the 
visibility of the CAST processes was due to (or could be enhanced by) a
level-dependent asymmetry of tunneling processes. This would lead, however, also to
an asymmetry in the gate voltage dependence, e.g. the CAST feature would be enhanced 
on the right side of the Coulomb diamond, but reduced on the left.  
When including relaxation processes of the excited states 
(not shown) we find that weak relaxation can at first enhance
the CAST feature, while strong relaxation (with rate many times larger 
than $\Gamma$) suppresses the CAST feature such that it becomes invisible in the differential conductance. As has been discussed in recent work~\cite{pfannkuche}
such a  suppression of CAST features could occur in many experimental situations 
due to relaxation processes involving phonons, photons or nuclear spins.
However, we find that even under the influence of a relatively strong relaxation  the Fano factor still displays the CAST processes, demonstrating once again the importance of shot noise measurements for a full characterization of the underlying transport mechanisms.
The Fano factor displayed in the right panel of Fig.~\ref{fig:didv_Fano_triangle} shows the
inelastic co-tunneling excitation energy scale at $e V_{\rm co}=0.5\,{\rm meV}$ 
when it becomes super-Poissonian. For higher bias, the CAST angular shape inside the Coulomb diamond is very notable as the Fano factor reaches a yet higher value of
$F\approx 2$, due to the additional competition of CAST processes 
with the already present inelastic co-tunneling processes.

\begin{figure}
\centerline{\includegraphics[width=12cm]{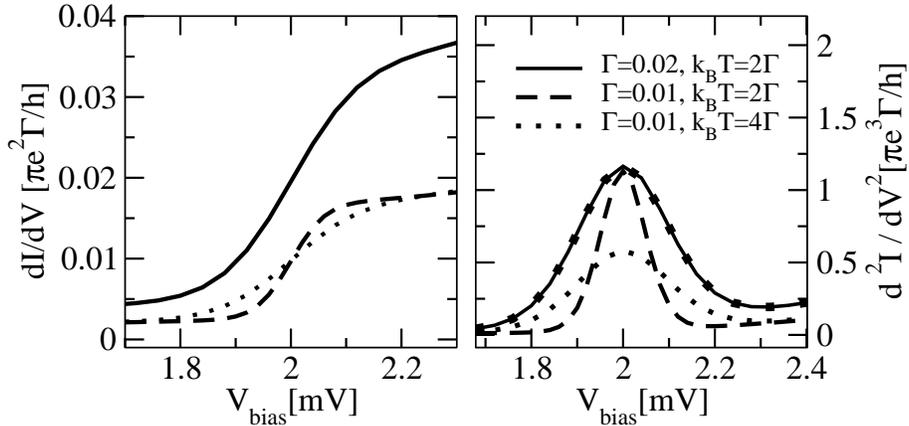}}
\caption{{\it Left panel}: Differential conductance $dI/dV$
   vs. bias voltage for the same parameters as in Fig.~\ref{fig:didv_Fano} 
   at fixed gate voltage $V_{\rm gate}=0$. The coupling $\Gamma$ and the temperature $T$ are 
   varied as indicated on the right panel. {\it Right panel}: Second derivative of the current, $ d^2I/dV^2$,
vs. bias voltage for the same parameters. The width of the peak scales linearly with temperature.}
 \label{fig:gamma_variation_new}
\end{figure}

{\bf Analysis of CAST transport features.}
We now address the question how the described co-tunneling and CAST transport features depend 
on the coupling strength and temperature, by example of the differential conductance for the parameters
of Fig.~\ref{fig:didv_Fano}. The left panel of Fig.~\ref{fig:gamma_variation_new}
displays a step in the differential conductance at the
inelastic co-tunneling excitation energy. An analysis of the step height for different
coupling strengths shows a linear dependence on the total coupling $\Gamma$. Since we have normalized the current and hence also the differential conductance by a factor $\Gamma$
this means that the co-tunneling shoulder height scales proportional to
$\Gamma^2$, as anticipated for a second order transport process. 
The second derivative of the current depicted in the right panel of
Fig.~\ref{fig:gamma_variation_new} further elaborates this fact. Here, the dotted curve for the weaker coupling $\Gamma$, when multiplied by the factor $\Gamma_{\rm solid}/\Gamma_{\rm dotted}=2$, practically coincides (fat dots) with the solid curve of a higher $\Gamma$. 
The absolute height of the co-tunneling feature is further influenced by the value of the inelastic
co-tunneling excitation energy $\epsilon_{\rm co}$ itself, 
as well as the energy $\epsilon_{\rm seq}(V_{\rm gate})$. 
There is no simple relation for the gate-voltage dependence of the step height. This is because the second order transition rates depend on the energy differences between the initial/final states of the transition and the virtually occupied intermediate states, and there are several possible virtual states, namely the empty dot state and six doubly occupied states. The step height is lowest in the center of the Coulomb diamond, as can be seen in the left panel of Fig.~\ref{fig:didv_Fano}.

To analyze the width of the inelastic co-tunneling feature we compare the two
curves (dashed and dotted) of the left panel of Fig.~\ref{fig:gamma_variation_new}, where both curves correspond to the same total coupling $\Gamma$ but to different temperatures. An obvious temperature broadening is observed. Considering the second derivative of the current for different temperatures we note that the width scales linearly with temperature, with full width at half maximum of $\sim 7-8\, k_{\rm B} T$. Though our theory can not reach the low temperature regime $k_{\rm B} T \ll \Gamma$, as mentioned above, the line width broadening is clearly observed at the conductance peak at the sequential threshold already for the value $k_{\rm B} T = 2 \Gamma$ used in the present analysis.
No such broadening is observed at the inelastic co-tunneling feature, {\it even though} CAST processes are also involved for the considered regime~\cite{footnote}.
This result does not support the interpretation of experimental findings provided in Ref.~\cite{franceschi}. There, a large broadening of a transport feature with CAST processes
was interpreted as line width broadening of size $\Gamma_{\rm R}$. However, the temperature dependence 
was not studied in that work. It remains open, whether there is truly a conflict of results, or
merely an experimental broadening of unknown physical origin, which is not considered in our theory.

{\bf Summary.}
In summary we discussed the influence of co-tunneling and co-tunneling
assisted sequential tunneling on the differential conductance and
 shot noise (Fano factor) inside the Coulomb blockade regime. We found that
the width of the features in the differential conductance  due to inelastic
co-tunneling is reduced to the thermal width, even when CAST processes are possible,
whereas the height of the feature scales with $\Gamma^2$. Additionally, we have studied a situation
in which the energy scale for CAST processes is separated from inelastic co-tunneling.
The onset of CAST processes varies with gate voltage parallel to the Coulomb diamond edge, leading to an ``angular'' shape inside the Coulomb diamond.
The Fano factor, already super-Poissonian due to inelastic co-tunneling processes, is even further enhanced by the CAST processes. While the conductance features due to CAST processes are suppressed by relaxation processes of excited states, the super-Poissonian Fano factor is much more robust against this decay.

{\em Acknowledgments.}
We acknowledge stimulating discussions with Maarten Wegewijs, Martin Leijnse,
J\"urgen K\"onig, Jonas Pedersen, and Andreas Wacker.

\newpage

\begin{thebibliography}{99}
\bibitem{footnote} The width scales with temperature also if only inelastic co-tunneling processes and no
additional CAST processes are possible at $V_{\rm co}$.
\end{thebibliography}
\end{document}